\documentstyle[prl,twocolumn,aps,epsfig]{revtex}
\def\prn#1{{\left(#1\right)}}

\def\fig_width{3. in} 

\begin{document}
\title{Solidification pipes: from solder pots to igneous rocks}
\author{M. Stewart Siu$^{a}$ and Dmitry Budker$^{b,c}$\cite{byline}}
\address {$^a$Department of
Applied and Engineering Physics, Cornell University, Ithaca NY
14850
\\ $^b$Department of Physics, University of California,
Berkeley, CA 94720-7300
\\ $^c$Nuclear Science Division, Lawrence Berkeley National Laboratory, Berkeley CA
94720}
\date{\today}
\maketitle

\begin{abstract}

When a substance that shrinks in volume as it solidifies (for
example, lead) is melted in a container and then cooled, a deep
hole is often found in the center after resolidification. We use a
simple model to describe the shape of the pipe and compare it with
experimental results.
\end{abstract}
\pacs{PACS. 81.30.Fb} In an experiment that involves atomic beams
of thallium \cite{Budker}, it was noticed that a deep narrow hole
was formed in the thallium that melted and resolidified. The hole
that formed was at the center of the container and extended from
the surface to nearly the bottom. It was surmised that the
phenomenon was due to the change in volume of thallium during
solidification. Such formation is sometimes known as ``pipe" in
metallurgy \cite{Chalmers}. In this note, we discuss a simple
model of pipe formation and compare it with straightforward
experiments that can be carried out in classrooms.

Suppose a molten substance is cooling in a circular cylinder.
Assuming that solidification occurs from the side walls of the
container inwards in the radial direction and neglecting the
surface tension effects, we should expect the liquid level to drop
as a layer of solid is formed because of the higher density of the
solid. Consider a newly solidified layer of thickness $dr$. Let
$\rho_{s}$ and $\rho_{l}$ be the solid and liquid densities
respectively, and let $h(r)$ be the height of solid as a function
of radius $r$. Equating the mass before and after solidification,
one obtains a differential equation:

\begin{equation}
\pi r^2 h\rho_{l}=\pi\prn{r-dr}^2\prn{h-dh}\rho_{l}+2\pi
 rhdr\rho_{s}.
\end{equation}
\newline Keeping only first order differentials, we get:

\begin{equation}
 \frac{dh}{h}=2\prn{\frac{\rho_{s}-\rho_{l}}{\rho_{l}}}\frac{dr}{r}.
\end{equation}
\newline With the boundary condition of $h(R)=h_{0}$, where $R$ and
$h_{0}$ are the radius of the container and the initial liquid
level respectively, the solution is:

\begin{equation}
 h=h_{0}\prn{r \over R}^{2\alpha},
 ~\alpha=\frac{\rho_{s}-\rho_{l}}{\rho_{l}}\geq0.
\end{equation}

This solution (plotted in Fig. 1 for the parameters of an
experiment described below) gives a sharp hole in the center, the
shape of which, for a given container and liquid volume, is
determined by $\alpha$, the fractional density change.

With this simple model in mind, we have performed solidification
experiments with various substances (this time omitting the highly
toxic thallium). The changes in densities upon solidification for
these materials and for thallium are listed in Table I
\cite{Chalmers,Metal,Merck}. As expected, pipes are observed in
all materials tested except Wood's metal (an alloy of 50\% Bi,
25\% Pb, 12.5\% Cd and 12.5\% Sn). Indeed, Wood's metal has the
property that the volume changes little during solidification.
Note that for substances that expand upon solidification (water,
bismuth, antimony and gallium), no "anti-pipe" is formed because
the liquid is pushed out by the expanded solidified material and
assumes a horizontal level.

Photographs of several experimental samples are shown in Figures
2-5. Figure 2 shows a sample of conventional solder alloy (60\%
lead, 40\% tin) that was melted and poured into a glass beaker
where it cooled and solidified. The sample was then cut through
the center of the pipe, the resulting cross-section is shown in
Figure 3.

Comparing the shape of the pipe predicted by our simple model
(Fig. 1) to the one observed experimentally (Figs. 2 and 3), one
finds that, while the shape is reproduced qualitatively, there are
also significant discrepancies. First, the pipe does not actually
go to the bottom of the container as the model predicts. Second,
the pipe in the experiment turns out to be much wider. Presumably
this is because we have assumed that solidification occurs only
from the sides (see below).

In fact, when cooling from the surface and the bottom becomes
significant, other scenarios in addition to pipe formation are
possible. Fig. 4 shows a solidified lead sample, in which a layer
of solid on the surface covers the pipe, turning it into a cavity.
We can see that the cavity width is greater than the pipe width
predicted from Equation 3. Qualitatively this can be understood
from the requirement of mass conservation: the material solidified
on the top does not have a chance to fill the pipe.

To reduce the relative solidification rate from the surface, we
attempted accelerated cooling from the sides by putting a beaker
with molten solder into a water bath. This time, instead of a deep
pipe, a surface recession shown in Fig. 5 was observed. To explain
this observation, we modified the model by adding a term to
account for solidification from the bottom.

Let $k$ be the ratio of the solidification rate of the bottom to
that of the sides. In order to keep the model as simple as
possible, we assume $k=h_{r=0}/R$. (Note that this would not be a
valid approximation for large $k$. If the solidification from the
bottom is sufficiently rapid, the entire substance solidifies
before solidification from the sides reaches $r=0$. In the cases
discussed here, however, the liquid level is high and the cooling
rate from the bottom is about the same as that from the sides, so
the assumption can be safely granted.) The differential equation
analogous to Equation 1, with the shorthand $h'=h-k(R-r)$, is
then:

\begin{eqnarray}
\lefteqn{\pi r^2h' \rho_{l}=\pi\prn{r-dr}^2\prn{h'-kdr-dh}\rho_{l}
\nonumber } \hspace{1.5cm} \\ && +2\pi
rh'dr\rho_{s}+\pi\prn{r-dr}^2 kdr\rho_{s}.
\end{eqnarray}
Simplifying, we get
\begin{equation}
 \frac{dh}{dr} = \frac{2\alpha\prn{h-kR}}{r} + 3k\alpha .
\end{equation}
\newline The solution is a long algebraic expression, which we
omit here, but the solution plot (for $k=1$) is given in Fig. 6.
Comparing it to the picture of the sample (Fig. 5), one can find
close resemblance between the two.

So far we have neglected the effect of surface tension (a simple
discussion of surface tension is given in \cite{Kikoin}, for
example). If wetting occurs at the solid-liquid interface of the
solidifying substance, the surface of the liquid will not be flat,
and the curvature of the surface will affect the final shape of
the solid. However, it is reasonable to assume that this effect
only becomes significant when the dimension of the contained
liquid is "capillary" --- i.e., the radius of curvature of the
surface near the wall, $a$, becomes comparable to the radius of
the liquid surface, $r$. From dimensional analysis, we expect
$a^2\sim \frac{\sigma}{\rho g}$. Plugging in realistic parameters,
for example, $\rho_{l}= 10^4 kg/m^3$(for metal), $\sigma=0.5 N/m$,
we obtain $a\sim 2 mm$. This means that surface tension only
becomes important near the center of the container. The effect
should be observable at the bottom of the pipe. Qualitatively, we
would expect the bottom to be more concave than predicted by our
model due to the curved liquid surface, and this is indeed the
case (see Fig. 3).

In conclusion, we have discussed the mechanism of formation of
surface pipes upon resolidification of materials with
$\rho_{l}/\rho_{s}<1$. These prominent formations can often be
observed in solder pots, candle containers, etc. They are
important in metallurgy \cite{Chalmers} where they have to be
taken into account in casting processes. Similar formations also
occur in igneous rocks due to density changes of magma on
solidification \cite{Grout}. However, it is often difficult to
separate this effect from a large number of other factors that
determine the structure and texture of igneous rocks.

The authors are grateful to D. E. Brown, D. DeMille, J. Demouthe,
D. F. Kimball, S. M. Rochester, V. V. Yashchuk for useful
discussions. This work was supported by National Science
Foundation under CAREER Grant No. PHY-9733479.

\begin{figure}
\centering \epsfig{file=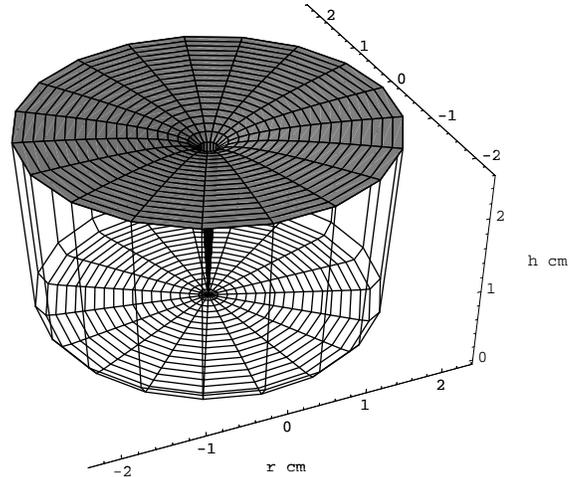,width=3.2in} \caption{A plot
of solution (3) with $h_{0}=2.5~cm,$ $R=2.3~cm,~\alpha=0.025$.}
\label{Fig. 1}
\end{figure}

\begin{figure} \centering
\epsfig{file=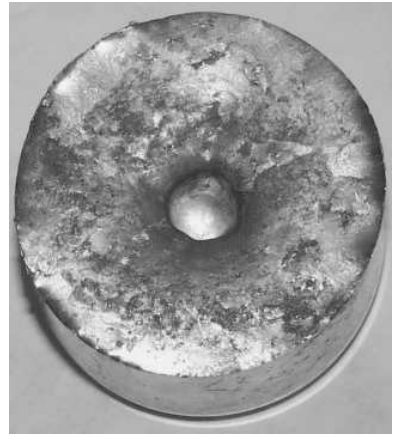,width=2in} 
\vspace{2 mm} \caption{Top view of the solder sample.
$h_{0}\approx 2.5~cm,$ $R\approx 2.3~cm$. } \label{Fig. 2}
\end{figure}

\begin{figure}
\centering
\epsfig{file=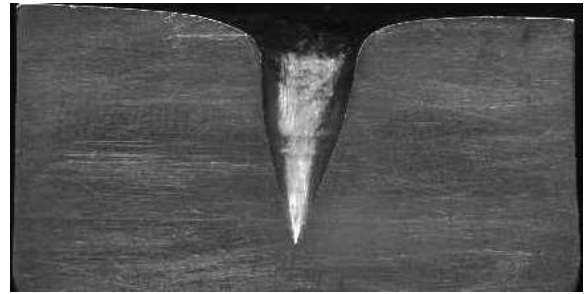,width=3in} 
\vspace{5 mm} \caption{Cross section of the solder sample in
Fig.2.} \label{Fig. 3}
\end{figure}

\begin{center}
\begin{figure}
\epsfig{file=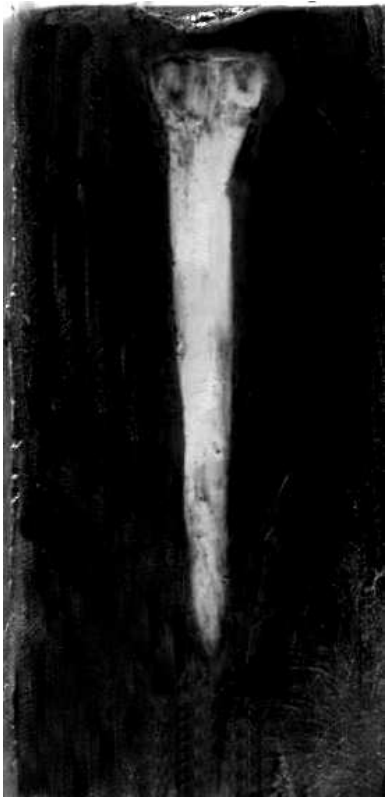,width=2in} 
\vspace{5 mm} \caption{Cross section of the lead sample with
$h_{0}\approx 9~cm,$ $R\approx 2.2~cm$. Note that the pipe is
closed from the top, forming a cavity.} \label{Fig. 4}
\end{figure}
\end{center}

\begin{figure}
\epsfig{file=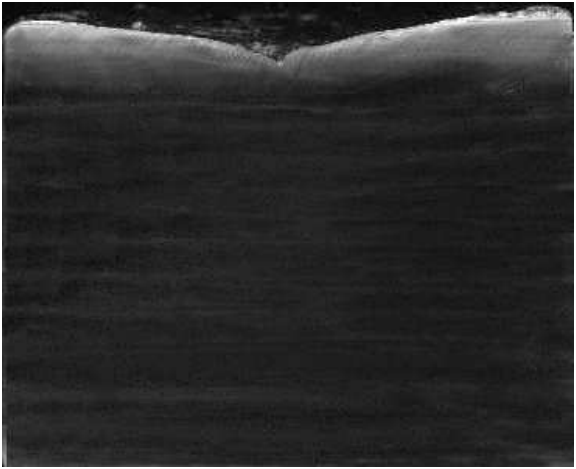,width=3in} 
\vspace{5 mm} \caption{Solder sample cooled in a water bath.
$h_{0}\approx 3.6~cm$, $h_{0}\approx 2.3~cm$. The curvature on the
sides is, presumably, due to anti-wetting of solder with the glass
surface of the container.} \label{Fig. 5}
\end{figure}

\begin{figure}
\epsfig{file=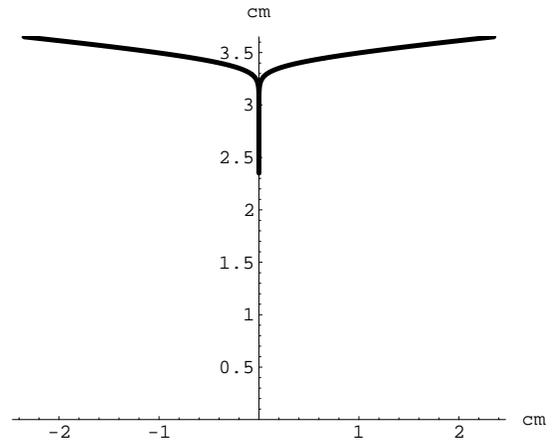, width=3in} \vspace{5 mm} \caption{A plot
of the solution for Equation (5) with $k=1$ and other parameters
as those for the sample in Fig. 5. An extremely narrow pipe
(radius $< 10^{-3}$ cm) is present in the plot, but as one would
reasonably expect, such delicate structure is not found in the
sample.} \label{Fig. 6}
\end{figure}

\end{document}